\documentstyle[12pt]{article}

  \let\LARGE=\large
 \let\large=\normalsize
\newcommand{\be}{\begin{equation}}
\newcommand{\ee}{\end{equation}}
\newcommand{\ba}{\begin{array}{c}}
\newcommand{\ea}{\end{array}}
\newcommand{\dis}{\displaystyle}

\begin{document}

\begin{titlepage} \vspace{0.2in} \begin{flushright}
CPT-92/P.2795\\ \end{flushright} \vspace*{1.5cm}
\begin{center} {\LARGE \bf  Rare Kaon Decays in  the
$1/N_c$-Expansion\\} \vspace*{0.8cm} {\bf C. Bruno and J.
Prades}\\ \vspace*{1cm} Centre de Physique Th\'eorique,
C.N.R.S. - Luminy, Case 907 \\ F-13288 Marseille Cedex 9,
France \\ \vspace*{1.8cm} {\bf   Abstract  \\ } \end{center}
\indent

We study the unknown coupling constants that appear at ${\cal O}
(p^4)$ in the Chiral Perturbation Theory analysis of $K \to \pi
\gamma^* \to \pi l^+ l^-$, $K^{+-} \to \pi^{+-} \gamma \gamma$
\cite{epr1} and $K \to \pi \pi \gamma$ \cite{ep} decays. To that end,
we compute the chiral realization of the $\Delta S \, = \, 1$
Hamiltonian in the framework of the $1/N_c$-expansion of the
low-energy action proposed in Ref. \cite{pr}.  The phenomenological
implications are also discussed. \vfill \begin{flushleft}
CPT-92/P.2795\\
 July 1992 \end{flushleft} \end{titlepage}

\section{Introduction} \label{1} \indent

In this work we shall study the unknown coupling
constants that govern the analysis of $K \to \pi \gamma^*
\to \pi l^+ l^-$, $K^{+-} \to \pi^{+-} \gamma \gamma$
\cite{epr1} and $K \to \pi \pi \gamma$ \cite{ep} decays at
${\cal O} (p^4)$ in the framework of Chiral Perturbation
 Theory (ChPT).  We shall restrict ourselves to the
non-anomalous sector of the theory. (For a recent
discussion of anomalous non-leptonic decays see Ref.
\cite{ep}.) These transitions have become particularly
interesting because of possible CP-violation effects in
rare $K$-decays. The various tests on this subject
discussed in Ref. \cite{epr1}  depend on the values of
these coupling constants. These are not fixed  by chiral
symmetry requirements alone but are, in principle,
determined by the dynamics of the underlying theory.

Recently there have been attempts to derive the low
energy effective   chiral action of QCD \cite{ert,bbr} as
well as some of the coupling  constants of the effective chiral
Lagrangian of  strangeness-changing four-quark Hamiltonian \cite{pr}.
Here we shall calculate
 the above mentioned couplings in the framework of this
approach.

It is well known that gauge invariance together with chiral
symmetry forbid the ${\cal O} (p^2)$ contribution
\cite{epr1} to $K$-decays with at most one  pion in the
final state. The first non-vanishing contributions to the
transition amplitudes $K \to \pi \gamma^*$ and $K \to \pi
\gamma \gamma$  appear at ${\cal O} (p^4)$. The
contributions from chiral loops have been calculated in
Ref. \cite{epr1}. The $K \to \pi \pi \gamma$ decays have
been analysed in Ref. \cite{ep}  to the same order in the
chiral expansion. In Refs. \cite{epr1,gef,epr2}, some of the
couplings constants we are interested in have been
calculated within  the context of various models.

The effective non-leptonic chiral Lagrangian consistent
with the octet structure of the dominant $\Delta S \, = \,
1$ weak Hamiltonian has been classified
 in Refs. \cite{gef,kam}. This Lagrangian, when restricted
to the rare kaon decays processes we are interested in
can be parametrized in
 terms of the set of coupling constants $\omega_1$,
$\omega_2$, $\omega_4$,  $\omega_1'$ and $\omega_2'$,
introduced in Refs. \cite{epr1,ep} as follows,

\be \label{effla} \renewcommand{\arraystretch}{1.3}
\begin{array}{l} {\cal L}^{(4)}_{eff} \, = \, - \frac{\dis
G_F}{\dis \sqrt 2} \, V_{ud} V^*_{us} \, \frac{\dis
f_\pi^2}{\dis 4} \, g_8 \,  \left[ i \, \omega_1 \langle
f_{(+)}^{\mu \nu} \left\{ \Delta_{3 2} \, , \, \xi_\mu \xi_\nu
\right\} \rangle \, + \,  2 \, i \, \omega_2 \langle
f_{(+)}^{\mu \nu} \xi_\mu \Delta_{3 2} \xi_\nu  \rangle
\right. \\ \left.  \hspace*{0.5cm} + \, \frac{\dis 4}{\dis 3} \,
\omega_4^{- -} \,  \langle f_{(-)}^{\mu \nu} f_{(-) \mu \nu}
\Delta_{3 2} \rangle \, -  \, 4 \, \omega_4^{+ +}  \, \langle
f_{(+)}^{\mu \nu} f_{(+) \mu \nu} \Delta_{3 2} \rangle  - \, 2
\, \omega_4^{- +} \, \langle \left\{f_{(-)}^{\mu \nu} \, , \,
f_{(+) \mu \nu} \right\} \Delta_{3 2} \rangle \right. \\ \left.
\hspace*{2.5cm}  + \, i \, \omega '_1 \langle f_{(-)}^{\mu
\nu}  \left\{ \Delta_{3 2} \, , \, \xi_\mu \xi_\nu \right\}
\rangle  \, + \, 2 \, i \, \omega '_2 \langle f_{(-)}^{\mu \nu}
\xi_\mu  \Delta_{3 2} \xi_\nu \rangle \right] \, + \, {\rm
h.c.} \\
 \hspace*{5cm} + \, {\rm non-octet \, \, operators}, \ea
\renewcommand{\arraystretch}{1} \ee

\noindent with $\omega_4 \, = \, \omega_4^{- -} \, + \,
\omega_4^{+ +} \, + \,  \omega_4^{- +}$, where  $\langle \,
\, \rangle$ denotes a trace in flavour. The notation here is defined
below:

\be \renewcommand{\arraystretch}{1.3} \ba f^{\mu
\nu}_{(\pm)} \, = \, \xi F^{\mu \nu}_L \xi^\dagger \pm
\xi^\dagger F^{\mu \nu}_R \xi  , \\ \Delta_{3 2} \, = \, \xi
\lambda_{3 2} \xi^\dagger , \\ \xi_\mu \, = \, i \,
\xi^\dagger D_\mu U \xi^\dagger \, = \, - i \,  \xi D_\mu
U^\dagger \xi , \ea \renewcommand{\arraystretch}{1} \ee

\noindent where $\lambda_{ab}$ is the $3 \times 3$
flavour matrix  $(\lambda_{ab})_{ij} \, = \, \delta_{ai}
\delta_{bj}$ and $\xi$ is chosen  such that

\be \ba U \, = \, \xi^2. \ea \ee

\noindent $U \equiv \exp \left( -\frac{\dis \sqrt 2 i
\Phi}{\dis f} \right)$ is a SU(3) matrix incorporating the
octet of pseudoscalar mesons,

\be \ba \Phi (x) \equiv \frac{\dis \vec{\lambda}}{\dis \sqrt
2} \, \vec{\varphi}
 = \, \left( \begin{array}{ccc} \frac{\dis \pi^0}{\dis \sqrt 2}
\, + \, \frac{\dis \eta}{\dis \sqrt 6}
 & \pi^+ & K^+ \\ \pi^- & - \frac{\dis \pi^0}{\dis \sqrt 2} \,
+ \, \frac{\dis \eta} {\dis \sqrt 6}    & K^0 \\ K^- & \bar K^0
& - \frac{\dis 2 \, \eta}{\dis \sqrt 6} \end{array}  \right) ,
\ea \ee

\noindent and $f \simeq f_\pi \, = \, 93.2$ MeV is the pion
decay constant at  lowest order. $D_\mu$ is a covariant
derivative \cite{ert} which acts on $U$. In the presence of
external electromagnetic fields only, it is defined as

\be D_\mu \, U \equiv  \partial_\mu \, U \, - \, i |e| \, A_\mu
[Q, U], \ee

\noindent where $e$ is the electron charge and $Q$ is a
diagonal $3 \times 3$ matrix that takes into account the
electromagnetic (u,d,s)-light-quark  charges, $Q\, = \,
\frac{\dis 1}{\dis 3} {\rm diag} (2, -1, -1)$. In this case we
also have

\be F^{\mu \nu}_L \, = \, F^{\mu \nu}_R \, = \, |e| Q F^{\mu
\nu}
 \, = \, |e| Q (\partial^\mu A^\nu  \, - \, \partial^\nu A^\mu).
\ee

\noindent In Eq. (\ref{effla}) $G_F$ is the Fermi constant
and $V_{ij}$ are the  Cabibbo-Kobayashi-Maskawa (CKM)
matrix elements. The $g_8$-coupling introduced in Eq.
(\ref{effla}) is the constant that modulates the octet
operator in the $\Delta S = 1$ effective Lagrangian of
order $p^2$; {\it i.e.},

\be \label{eg8} \ba {\cal L}^{(2)}_{eff} \, = \,  - \, \frac{\dis
G_F}{\dis \sqrt 2} \, V_{ud} \, V_{us}^* \, \, f_\pi^4 \,  \,
g_8 \, \langle \Delta_{3 2} \, \xi_\mu \xi^\mu \rangle \, +
\,  {\rm h.c.} \, + \, {\rm non-octet \, \, operators} . \ea \ee

We want to perform a calculation of the coupling
constants which appear in ${\cal L}_{eff}^{(4)}$ in  Eq.
(\ref{effla}) in the context of an expansion in powers of
$1/N_c$, ($N_c \, =$ number of colours). The paper is
organized as follows. In Section \ref{2b} we shall
introduce the $\Delta S \, = \, 1$ effective Hamiltonian. In
Section \ref{2} we shall study the large-$N_c$ limit
 result for the counterterms defined in Eq. (\ref{effla}). In
Section  \ref{4}, the effective realization of the $\Delta S
\, = \, 1$ four-fermionic  Hamiltonian is studied including
terms of ${\cal O} (N_c(\alpha_s N_c))$ as in Ref. \cite{pr}.
Phenomenological implications of our results are presented in Section
\ref{5} and finally, in Section \ref{6}, the conclusions are
given.

\section{The $\Delta S \, = \, 1$ effective Hamiltonian}
\label{2b} \indent

The $\Delta S \, = \, 1$ chiral Lagrangians in Eq.
(\ref{effla}) and  (\ref{eg8}) are part of the effective
realization of the corresponding  Standard Model (SM)
sector in terms of low-energy degrees of freedom. In the
SM with three flavours, once the heaviest particles
(W-boson, t-, b- and c-quarks) have been integrated out,
the complete basis of operators of  weak and strong
origin that induce strangeness changing processes with
$\Delta S \, = \, 1$ {\it via} W-exchange is given by,

\be \label{qoper} \renewcommand{\arraystretch}{1.3}
\begin{array}{l} Q_1 \, = \, 4 (\bar s_L \gamma^\mu d_L)
(\bar u_L \gamma_\mu u_L) \\ Q_2 \, = \, 4 (\bar s_L
\gamma^\mu u_L) (\bar u_L \gamma_\mu d_L) \\ Q_3 \, =
\, 4 (\bar s_L \gamma^\mu d_L) {\dis \sum_{q=u,d,s}} (\bar
q_L \gamma_\mu q_L) \\ Q_4 \, = \, 4 {\dis
\sum_{q=u,d,s}} (\bar s_L \gamma^\mu q_L)  (\bar q_L
\gamma_\mu d_L) \\ Q_5 \, = \, 4 (\bar s_L \gamma^\mu
d_L) {\dis \sum_{q=u,d,s}}  (\bar q_R \gamma_\mu q_R) \\
Q_6 \, = \, - 8 {\dis \sum_{q=u,d,s}}  (\bar s_L q_R) (\bar
q_R d_L). \ea \renewcommand{\arraystretch}{1} \ee

\noindent where $\Psi^R_L \equiv \frac{\dis 1}{\dis 2} \, (
1 \pm \gamma_5) \,  \Psi(x)$, and summation over colour
indices is understood inside each bracket.

The reduction of the SM Lagrangian to an effective
electroweak Hamiltonian has been discussed in the
literature \cite{sm}.   The structure of the effective
non-leptonic low-energy Hamiltonian is  the following,

\be \label{hamil} \renewcommand{\arraystretch}{1.3}
\begin{array}{l} {\cal H}^{\Delta S = 1}_{eff} \, =   \frac{\dis
G_F}{\dis \sqrt 2} \,  V_{ud} V^*_{us} \, \left\{ \frac{\dis
1}{\dis 2} C_+ (\mu^2) \,  (Q_2 \, + \, Q_1) \, + \, \frac{\dis
1}{\dis 2} C_- (\mu^2) \,  (Q_2 \, - \, Q_1) \right. \\ \left.  +
\, C_3 (\mu^2) \, Q_3 \, + \, C_4 (\mu^2) \, (Q_3 \, + \, Q_2
\, - \, Q_1)  \, + \, C_5 (\mu^2) \, Q_5 \, + \, C_6 (\mu^2) \,
Q_6 \right\} \, + \,
 {\rm h.c.} \ea \renewcommand{\arraystretch}{1} \ee

\noindent The Wilson coefficients $C_\pm$ and $C_i$, $i =
3,4,5,6$ are known functions of the heavy masses and the
renormalization scale $\mu$ beyond the leading
logarithmic approximation \cite{buras}. In the limit we are
working here we shall use these coefficients in the
leading logarithmic
 approximation. Of course, the matrix elements of the
$Q_{i = 1,\cdots,6}$ operators must depend on the $\mu$
scale in such a way that physical amplitudes do not
depend on it. We shall be only interested in the octet
components of the four-quark operators that induce
$\Delta I = 1/2$ transitions (due to the octet dominance
 of the $\Delta I = 1/2$ enhancement); namely, $Q_- \equiv
Q_2 - Q_1$, the  octet part of $Q_+ \equiv Q_2 + Q_1$ and
$Q_3, \, Q_4, \, Q_5, \, Q_6$ which  are pure octet
operators.

To these $\Delta S = 1$ operators in (\ref{qoper}) of weak
and strong origin we have to add two more operators that
induce  $\Delta S \, = \, 1$ transitions, coming from the
so-called  electroweak penguins \cite{gw},

\be \label{q7} \ba Q_7^V \, = \, \frac{\dis e^2}{\dis 2 \pi} \,
( \bar s_L \gamma^\mu d_L ) \, \,  \bar l \gamma_\mu l
\hspace*{1cm} {\rm and} \hspace*{1cm} Q_7^A \, = \,
\frac{\dis e^2}{\dis 2 \pi} \, ( \bar s_L \gamma^\mu d_L ) \,
\,  \bar l \gamma_\mu \gamma_5 l,  \ea \ee

\vspace*{0.5cm}

\noindent with their corresponding Wilson coefficients,
$C_7^V (\mu^2)$ and  $C_7^A (\mu^2)$. So the matrix
elements of the $Q_{i = 1, \cdots, 6}$ operators have to be
evaluated to order $\alpha\, =\, e^2/4 \pi$ whereas  the
matrix element  of $Q_7^{A,V}$ must be calculated to
order $\alpha^0$.  The effective Hamiltonian for
$Q_7^{A,V}$ can be written as

\be \label{gwh} \begin{array}{l} {\cal H}^{Q_7}_{eff} \, = \,
\frac{\dis G_F}{\dis \sqrt 2} \, V_{ud} V^*_{us} \, \left\{
C_7^V (\mu^2) \, Q_7^V \, + \, C_7^A (\mu^2) \,  Q_7^A
\right\} \, + \, {\rm h.c.} \ea \ee

\vspace*{0.5cm}

\noindent The Wilson coefficient $C_7^A$ only receives
contributions from the so-called $Z$ penguin and $W$ box
diagrams. In the present paper we are just interested in
transitions that are mediated by external photons,
therefore we are not going to consider the electroweak
penguin operator modulated by $C_7^A$ and we have
$C_7^V \, = \, C_7^\gamma$. The expression for the
electromagnetic penguin Wilson coefficient
$C_7^\gamma (\mu^2)$ can be found in Ref. \cite{gw}. This
Wilson coefficient  contains a complex phase coming from
the CKM matrix elements that can give rise to ``direct''
CP-violation effects. In the rest of the paper when we
refer to the $Q_7$ operator we would mean
$Q_7^\gamma$.

The other two $\Delta S \, = \, 1$ operators which arise
from electromagnetic  penguin-like diagrams, see Ref.
\cite{bw}, start to contribute at order  $\alpha^2$.

\section{The $\Delta S \, = \, 1$ effective Hamiltonian in
the large-$N_c$  limit} \label{2} \indent

In the large-$N_c$ limit, the $\Delta \, S \,=\, 1$  effective
Hamiltonian is given by,

\be \label{hanc} \renewcommand{\arraystretch}{1.3}
\begin{array}{l} {\cal H}^{\Delta S = 1}_{eff} \, =   \frac{\dis
G_F}{\dis \sqrt 2} \,  V_{ud} V^*_{us} \, Q_2 \, + \, {\rm h.c.} .
\ea \renewcommand{\arraystretch}{1} \ee

\noindent In this limit we just need to calculate the
effective action for  the factorizable pattern of the
four-quark operator $Q_2$. This can be
 performed in a model independent way by doing
appropriate products of the low-energy realization of
quark currents in the framework of   effective chiral
Lagrangians.

To obtain the large-$N_c$ limit effective realization at
${\cal O}(p^4)$ we need to know the effective realization
of quark currents up to ${\cal O}(p^3)$.  These can be
easily derived from the ${\cal O} (p^2)$ and ${\cal O}
(p^4)$  strong effective chiral Lagrangian given in Refs.
\cite{gl1,sw} in the
 presence of external sources. In addition, we want to
keep only the octet  component of the isospin-$1/2$
operators. In this limit the value of the  $g_8$-coupling in Eq.
(\ref{eg8}) is $(g_8)_{1/N_c} \, = \, 3/5$, to be compared with the
experimental value from $K \to \pi \pi$ decays,
$|g_8|_{\rm exp} \simeq 5.1$. For the coupling constants
$\omega_{1,2,4}$ and $\omega_{1,2}'$ of the ${\cal O}
(p^4)$ effective Lagrangian in (\ref{effla}) we then obtain
the  following results,

\be \label{w1w2} \renewcommand{\arraystretch}{1.3} \ba
\omega_1 \, = \, \omega_2 , \\ g_8 \, \omega_2 \, = \, 8 \,
(g_8)_{1/N_c} \, L_9, \\ g_8 \, \omega_4 \, = \, 12 \,
(g_8)_{1/N_c} \, L_{10},\\ g_8 \, \omega'_1 \, = \, 8 \,
(g_8)_{1/N_c} \, (L_9 \, - \, 2 \, L_{10}), \\ {\rm and}\\
\omega'_2 \, = \, 0. \ea \renewcommand{\arraystretch}{1}
\ee

\vspace*{0.5cm}

\noindent Here $L_9$ and $L_{10}$ are two of the couplings
that appear modulating local terms in the chiral
Lagrangian at order $p^4$. In the notation of  Gasser and
Leutwyler \cite{gl1}, they read as follows,

\be \label{lem} - i \, L_9 \, \langle F^{\mu \nu}_L D_\mu U
D_\nu U^\dagger \, +  \, F^{\mu \nu}_R D_\mu U^\dagger
D_\nu U \rangle \,+\, L_{10} \langle U F^{\mu \nu}_R
U^\dagger F_{L , \mu \nu} \rangle .  \ee

\vspace*{0.5cm}

The present experimental results on the $K^+ \to \pi^+
e^+ e^-$ process  \cite{bnl} allow for the determination of
the combination of coupling constants,

\be g_8 (\omega_1 \, + \, 2 \, \omega_2 )\, = \, 0.41 ^{+
0.10}_{- 0.05}. \ee

\noindent Numerically, in the large-$N_c$ limit we find

\be \label{lnc} g_8 (\omega_1 \, + \, 2 \, \omega_2 )\, = \,
24 \, (g_8)_{1/N_c} \, L_9 \, = \, 0.10 ; \ee

\noindent {\it i.e.}, a factor four lower than the
experimental value.

If one looks at Eq. (\ref{w1w2}), the naive approach would
be to consider  that $g_8$ factorizes in the r.h.s. to all
orders in the $1/N_c$-expansion  and therefore write
down the following result,

\be \label{w1w2c} \renewcommand{\arraystretch}{1.3} \ba
\omega_1 \, = \, \omega_2  \,  = \, 8 \, L_9, \hspace*{2cm}
\omega_4 \, = \, 12 \, L_{10}, \\ \omega'_1 \, = \, 8 \, (L_9
\, - \, 2 \, L_{10}) \hspace*{1cm} {\rm and} \hspace*{1cm}
\omega'_2 \, = \, 0. \ea \renewcommand{\arraystretch}{1}
\ee

\vspace*{0.5cm}

\noindent Then

\be \label{lnc2} \omega_1 \, + \, 2 \, \omega_2 \, = \, 24 \,
L_9 \, = \, 0.16 , \ee

\vspace*{0.5cm}

\noindent which is about twice the experimental value. In
Section \ref{4} we shall come back to the question of
whether or not  factorization of $g_8$ in Eq. (\ref{w1w2})
remains valid after including  the next-to-leading
corrections in the $1/N_c$-expansion.

It can be seen from the results above that we do not
obtain the octet  dominance relation $\omega_2 \, = \, 4 \,
L_9$ which was assumed in Ref.  \cite{epr1}. Our results
differ from those found in Ref. \cite{cheng}, where it is
claimed the numerical relation  $\omega_1 \, = \,
\omega_2 \, = \, 4 \, L_9$ in the large-$N_c$ limit.  In
addition we get $\omega_4 \, = \, 12 \, L_{10}$ which also
differs from Ref. \cite{cheng} where $\omega_4 \, = \, 0$.
The results found in Ref.  \cite{epr2} for $\omega_{1,2,4}$
from the ``weak deformation model'' are different to
those we find in the large-$N_c$ limit.

The amplitude for the process $K^{+-} \to \pi^{+-} \gamma
\gamma$ to lowest non-trivial order in ChPT was
calculated in Ref. \cite{epr1}. The result turns out to
depend on the following combination of coupling
constants which is  renormalization scale independent,

\be \label{hc} \renewcommand{\arraystretch}{1.3} \ba
\hat c \, = \, 32 \pi^2 \left[ 4 (L_9 \, + \, L_{10}) -
\frac{\dis 1}{\dis 3} (\omega_1 \, + \, 2 \, \omega_2 \, +  \,
2 \, \omega_4) \right] \\ \hspace*{2.5cm} = \, - \frac{\dis
32 \pi^2}{\dis 3} \, \left[ ( \omega_1 \, - \,  \omega_2 ) \, +
\, 3 \, ( \omega_2 \, - \, 4 L_9 ) \, + \, 2 \,  ( \omega_4 \, -
\, 6 \, L_{10}) \right]. \ea
\renewcommand{\arraystretch}{1} \ee

\vspace*{0.5cm}

\noindent The combination  $L_9 \, + \, L_{10}$ is a
renormalization scale invariant quantity that is
determined from the so-called structure term in $\pi \to
e \nu \gamma$ \cite{gl1,pdg} to be

\be \ba L_9 \, + \, L_{10}\, = \, (1.39 \pm 0.38) \times
10^{-3} . \ea \ee

\vspace*{0.5cm}

\noindent The combination $\omega_1 \, + \, 2 \,
\omega_2 \, + \, 2 \,  \omega_4$ is, of course, also
renormalization scale invariant.  In the large-$N_c$ limit
we find for the coupling constant $\hat c$  the following
value,

\be \label{clnc} \ba \hat c \, = \, - \, 128 \, \pi^2 \, \,
\frac{\dis (g_8)_{1/N_c}}{\dis g_8}  \, (L_9 \, + \, L_{10}) \,
= \, - \, 1.8 \, \, \frac{\dis (g_8)_{1/N_c}}{\dis g_8}. \ea \ee

\vspace*{0.5cm}

\noindent Again, if we assume that factorization of $g_8$
in the r.h.s. of
 Eq. (\ref{w1w2}) is valid, we would obtain

\be \label{clnc2} \ba \hat c \, = \, - \, 128 \, \pi^2 \, (L_9 \,
+ \, L_{10}) \, = \, - \, 1.8 . \ea \ee

\vspace*{0.5cm}

\noindent The ``weak deformation model'' of Ref.
\cite{epr2} predicts $\hat c \, = \, 0$.

In the  transition $K^+ \to \pi^+ \pi^0 \gamma$ at ${\cal O}
(p^4)$ there  appears the following combination of
coupling constants \cite{ep}:  $\omega_1 \, + \, 2 \,
\omega_2 \, - \, \omega '_1 \, + \, 2\, \omega ' _2$.  The
combinations $\omega_1 \, + \, 2 \, \omega_2 \, - \,
\omega '_1$ and  $\omega '_2$ are scale independent
separately. For these combinations, we get  the following
results in the large-$N_c$ limit:

\be \label{wpr} \renewcommand{\arraystretch}{1.3} \ba
g_8 \, (\omega_1 \, + \, 2 \, \omega_2 \, - \, \omega '_1) \,
=  \, 16 \, (g_8)_{1/N_c} \, (L_9 \, + \, L_{10}) \, = \, 0.01 \\
{\rm and} \\ \omega'_2 \, = \, 0. \ea
\renewcommand{\arraystretch}{1} \ee

\vspace*{0.5cm}

In the following Section  we shall estimate the
next-to-leading corrections in the $1/N_c$-expansion to
the couplings $g_8 \, \omega_{1,2,4}$ and $g_8 \,
\omega_{1,2}'$ in the  same spirit as the calculation done
for $g_8$ in Ref. \cite{pr}.

\section{The effective action of four-quark operators at
${\cal O}(N_c (\alpha_s N_c))$} \label{4} \indent

In order to calculate the counterterms we are interested
in, we shall
 need the effective realization of the $Q_{i = 1, \cdots, 7}$
operators at ${\cal O} (p^4)$. The procedure we shall
follow up is the one  described in Ref. \cite{pr} where the
corresponding calculation to ${\cal O} (p^2)$ has been
reported. As there, we shall work in the chiral limit; {\it
i.e.}, we neglect the (u,d,s)-light-quark masses.  Now, it is
the complete set of strangeness changing $\Delta S \, = \,
1$  operators $Q_{i=1,\cdots,7}$ that has to be taken into
account. Since  we want to do our analysis in the context
of the $1/N_c$-expansion, it is worth to recall the
behaviour in the large-$N_c$ limit of the various
parameters we have introduced: $f_\pi^2$, $L_9$, $L_{10}$,
$\omega_{1,2,4}$ and  $\omega '_{1,2}$ are order $N_c$;
$g_8$, $C_+$, $C_-$ and $C_7$ are order $1$ and $C_{3, 4, 5,
6}$ are order $1/N_c$. We want to perform a  calculation
of  the effective Hamiltonian in (\ref{hamil}) and
(\ref{gwh}) up to order  $N_c(\alpha_s N_c)$. Only the
$Q_+$, $Q_-$ and $Q_7$ operators are then needed at this
order, since all the other four-quark operators are
modulated  by Wilson coefficients that are already order
$1/N_c$. The  $Q_{i = 1, \cdots, 5}$ operators and $Q_{i=6,
7}$ operators have different  spinorial structure, thus  we
are going to study them separately. This will be done in
Section \ref{4.1}  and Section \ref{4.2}, respectively. We
shall summarize the corresponding  results for the
couplings $g_8 \, \omega_{1,2,4}$ and $g_8 \,
\omega'_{1,2}$ in Section \ref{4.3}.

\subsection{The effective action of the $Q_{i=1,\ldots,5}$
operators} \label{4.1} \indent

As it has been stated in the introduction of this Section,
we need the effective action of  $Q_-$ and $Q_+$ to order
$N_c (\alpha_s \, N_c)$; and to order $N_c^2$  for the
$Q_{3,4,5}$ operators. The results to ${\cal O} (p^4)$ for
the terms which are relevant to the decays $K \to \pi
\gamma^*$,  $K^{+-} \to \pi^{+-} \gamma \gamma$ and $K
\to \pi \pi \gamma$  are the following:

\be \label{eq-} \renewcommand{\arraystretch}{1.4}
\begin{array}{l} \langle Q_- \rangle \Rightarrow \\ - \,
f_\pi^2 \, \left\{ 2 \, i \, L_9 \,  \left(1 \, - \, \frac{\dis
g_1}{\dis N_c} \, - \, \gamma_-(\mu)\right)   \, \left[
\langle f^{\mu \nu}_{(+)} \,  \left\{ \Delta_{3 2} \, , \,
\xi_\mu \xi_\nu \right\} \rangle \, + \, 2 \, \langle f^{\mu
\nu}_{(+)}  \, \xi_\mu \Delta_{3 2} \xi_\nu \rangle \right]
\right. \\ \left.   + \, 4 \, L_{10} \, \left(1 \, - \, \frac{\dis
g_2}{\dis N_c}  \, - \, \gamma_-(\mu) \right) \,  \langle
f_{(-)}^{\mu \nu} \, f_{(-) \mu \nu} \, \Delta_{3 2} \rangle
\right. \\ \left.  + \, 2 \, i \, \left[ L_9 \,  \left(1 \, - \,
\frac{\dis g_1}{\dis N_c} \, - \, \gamma_-(\mu) \right)  \, -
\, 2 \, L_{10} \, \left(1 \, - \, \frac{\dis g_2}{\dis N_c}  \, -
\, \gamma_-(\mu) \right) \right] \, \langle f_{(-)}^{\mu
\nu} \, \left\{ \Delta_{3 2} \, , \,  \xi_\mu \xi_\nu \right\}
\rangle \right\} \\ + \, 4 \, \pi \, (2 \, H_1 \, + \, L_{10})
\left[ \frac{\dis \alpha_s}{\dis N_c} \, \left( \langle Q_6
\rangle \, + \,  \langle Q_4 \rangle \right) \, + \,  \frac{\dis
8}{\dis 3} \, \left( 1 \, - \, \frac{\dis g_3}{\dis N_c}  \, - \,
\gamma_-(\mu) \right) \, \langle Q_7 \rangle \right] ; \\
\ea \renewcommand{\arraystretch}{1} \ee

\vspace*{0.5cm}

\noindent and

\be \label{eq+} \renewcommand{\arraystretch}{1.4}
\begin{array}{l} \langle Q_+ \rangle \Rightarrow  \\ - \,
\frac{\dis f_\pi^2}{\dis 5} \, \left\{ 2\, i \, L_9  \,  \left(1 \,
+ \, \frac{\dis g_1}{\dis N_c} \, - \, \gamma_+(\mu) \right)
\,  \left[ \langle f^{\mu \nu}_{(+)} \,  \left\{ \Delta_{3 2} \, ,
\, \xi_\mu \xi_\nu \right\} \rangle \, + \, 2  \, \langle
f^{\mu \nu}_{(+)}  \, \xi_\mu \Delta_{3 2} \xi_\nu \rangle
\right] \right. \\ \left.  + \, 4 \, L_{10} \, \left(1 \, + \,
\frac{\dis g_2}{\dis N_c}  \, - \, \gamma_+(\mu) \right) \,
\langle f_{(-)}^{\mu \nu} \, f_{(-) \mu \nu} \,  \Delta_{3 2}
\rangle \right. \\ \left.  + \, 2 \, i \, \left[ L_9 \, \left(1 \, +
\, \frac{\dis g_1}{\dis N_c}  \, - \, \gamma_+(\mu) \right)
\, - \, 2 \, L_{10} \, \left(1 \, + \, \frac{\dis g_2}{\dis N_c}
\, - \, \gamma_+(\mu) \right) \right] \, \langle f_{(-)}^{\mu
\nu} \, \left\{ \Delta_{3 2} \, , \,  \xi_\mu \xi_\nu \right\}
\rangle \right\} \\  + \, 4 \, \pi \, (2 \, H_1 \, + \, L_{10}) \,
\left[ \frac{\dis \alpha_s}{\dis N_c} \,  \left( \langle Q_6
\rangle \, + \, \langle Q_4 \rangle \right) \, - \,  \frac{\dis
8}{\dis 3} \,  \left( 1 \, + \, \frac{\dis g_3}{\dis N_c} \, - \,
\gamma_+(\mu) \right) \, \langle Q_7 \rangle \right] \\ +
\, {\rm non-octet \, \, terms}.  \ea
\renewcommand{\arraystretch}{1} \ee

\vspace*{0.5cm}

\noindent The relevant terms for the effective action of
the $Q_3, \, Q_4, \, Q_5$  penguin operators are:

\be \label{eq3} \renewcommand{\arraystretch}{1.3}
\begin{array}{l} \langle Q_3 \rangle \Rightarrow {\cal O}
(N_c);  \ea \renewcommand{\arraystretch}{1} \ee

\vspace*{0.5cm}

\be \label{eq4} \renewcommand{\arraystretch}{1.3}
\begin{array}{l} \langle Q_4 \rangle \Rightarrow   - \,
f_\pi^2 \, \left\{ 2 \, i \, L_9 \, \left[ \langle f^{\mu
\nu}_{(+)} \,  \left\{ \Delta_{3 2} \, , \, \xi_\mu \xi_\nu
\right\} \rangle \, + \, 2 \, \langle f^{\mu \nu}_{(+)}  \,
\xi_\mu \Delta_{3 2} \xi_\nu \rangle \right] \right. \\ \left.
+ \, 4 \, L_{10} \, \langle f_{(-)}^{\mu \nu} \, f_{(-) \mu \nu}
\,  \Delta_{3 2} \rangle \, + \, 2 \, i \, \left(L_9 \, - \, 2 \,
L_{10} \right)  \, \langle f_{(-)}^{\mu \nu} \, \left\{
\Delta_{3 2} \, , \,  \xi_\mu \xi_\nu \right\} \rangle \right\}
\,;  \ea \renewcommand{\arraystretch}{1} \ee

\vspace*{0.4cm}

\be \label{eq5} \renewcommand{\arraystretch}{1.3}
\begin{array}{l} \hspace*{1cm} \langle Q_5 \rangle
\Rightarrow {\cal O} (N_c) .  \ea
\renewcommand{\arraystretch}{1} \ee

\vspace*{0.5cm}

In the expressions above there appear three coupling
constants of the chiral Lagrangian at order $p^4$
\cite{gl1}. Two of them, $L_9$ and $L_{10}$, have been
already introduced in Eq. (\ref{lem}). The coupling
constant $H_1$ is the constant that modulates a   contact
term between external sources in the chiral Lagrangian at
order $p^4$ \cite{gl1} as follows,

\be \label{h1} H_1 \, \langle F_R^{\mu \nu} \, F_{R , \mu
\nu} \, + \,  F_L^{\mu \nu} \, F_{L , \, \mu \nu} \rangle , \ee

\vspace*{0.5cm}

\noindent this constant is ${\cal O} (N_c)$ in the
large-$N_c$ limit. We have identified the coupling
constants $L_9$, $L_{10}$ and $H_1$ that appear in the
effective action of $Q_-$, $Q_+$ and $Q_4$ by comparing
our results with their respective expressions found in the
context of the mean-field approximation to the Nambu
Jona-Lasinio model  discussed in Ref. \cite{bbr}. Thus,
whenever their numerical values are  needed, we shall use
the values found in this model.

In Eqs. (\ref{eq-}) and (\ref{eq+}) we use the following
short-hand notation for the leading non-perturbative
gluonic corrections,

\be \label{gluco} \renewcommand{\arraystretch}{1.5}
\begin{array}{l} g_0 \, = \, 1 \, -   \, \frac{\dis 1}{\dis 2} \,\,
{\cal G} ; \\ g_1 \, = \, 1 \, - \, \frac{\dis 1}{\dis 6} \,
\frac{\dis f_\pi^2} {\dis 12 M_Q^2 L_9} \,\, {\cal G} ; \\ g_2
\, = \, 1 \, + \, \frac{\dis 1}{\dis 2} \, \frac{\dis f_\pi^2}
{\dis 24 M_Q^2 L_{10}} \,\, {\cal G} ; \\ g_3 \, = \, 1 \, + \,
\frac{\dis 1}{\dis 3} \, \frac{\dis f_\pi^2} {\dis 12 M_Q^2 (2
H_1 + L_{10})}\,\, {\cal G}, \ea
\renewcommand{\arraystretch}{1} \ee

\vspace*{0.5cm}

\noindent with

\be \label{cglu} {\cal G} \equiv \frac{\dis N_c \, \langle
\frac{\alpha_s}{\pi} G^2 \rangle} {\dis 16 \pi^2 f_\pi^4} ,
\ee

\noindent which is ${\cal O} (1)$ in the large-$N_c$ limit
and the constituent quark mass $M_Q$, which arises from
the following term,

\be \label{chb} \ba - M_Q (\bar q_R U q_L \, + \, \bar q_L
U^\dagger q_R ) . \ea \ee

\vspace*{0.5cm}

\noindent This term is equivalent to the mean-field
approximation of the Nambu  Jona-Lasinio mechanism
discussed in Ref. \cite{bbr}. There are also perturbative
gluonic corrections which we have gathered in the  terms,

\be \label{pert} \renewcommand{\arraystretch}{2}
\begin{array}{l} \gamma_- (\mu) \, = \, - \frac{\dis 3}{\dis
4} \, \frac{\dis \alpha_s} {\dis \pi} \, \ln \left( \frac{\dis
\mu^2}{\dis M_Q^2} \right) ;\\ \gamma_+ (\mu) \, = \,
\phantom{-} \frac{\dis 3}{\dis 4} \, \frac{\dis \alpha_s}
{\dis \pi} \, \ln \left( \frac{\dis \mu^2}{\dis M_Q^2} \right).
\ea \renewcommand{\arraystretch}{1} \ee

\vspace*{0.5cm}

\noindent {}From these expressions, it can be seen that
the anomalous dimensions of the effective action for the
$Q_-$ and $Q_+$ four-quark operators are the needed
ones to compensate the scale dependence of the Wilson
coefficients.

\subsection{The penguin operators $Q_6$ and $Q_7$}
\label{4.2} \indent

Let us first calculate the effective action of $Q_6$ since,
as discussed in the introduction of this Section, we only
need to know its result to leading ${\cal O}(N_c^2)$. The
calculation is rather straightforward and the result is

\be \label{eq6} \renewcommand{\arraystretch}{1.3}
\begin{array}{l} \langle Q_6 \rangle \Rightarrow  -
\frac{\dis \langle \overline \Psi \Psi \rangle} {\dis M_Q} \,
\frac{\dis N_c}{\dis 48 \pi^2} \, \left[ 3 \, i \, \langle f^{\mu
\nu}_{(+)} \,  \left\{ \Delta_{3 2} \, , \, \xi_\mu \xi_\nu
\right\} \rangle \, + \, 2 \, i  \, \langle f^{\mu \nu}_{(+)} \,
\xi_\mu \Delta_{3 2} \xi_\nu \rangle    \right. \\ \left.
\hspace*{0.5cm} + \, 3 \, \langle f_{(-)}^{\mu \nu} \, f_{(-)
\mu \nu} \, \Delta_{3 2}  \rangle \, + \, 2 \, \langle
f_{(+)}^{\mu \nu} \, f_{(+) \mu \nu} \,  \Delta_{3 2} \rangle
\right] .   \ea \renewcommand{\arraystretch}{1} \ee

\vspace*{0.5cm}

\noindent Here $\langle \overline \Psi \Psi \rangle$ is a
scale-dependent  quantity which, at one-loop level, is
defined by

\be \ba {\langle \overline \Psi \Psi \rangle}_{\mu^2} \, = \,
\left( \frac{\dis 1}{\dis 2} \, \ln \left( \frac{\dis \mu^2}
{\dis \Lambda_{\overline{MS}}^2}\right) \right)^{4/9} \,
\langle \widehat{\bar q q} \rangle . \ea \ee

\vspace*{0.5cm}

\noindent where $\langle \widehat{\bar q q} \rangle$ is
the scale invariant quark vacuum condensate. In the
large-$N_c$ limit the quark condensate is order $N_c$.

For the electromagnetic penguin operator $Q_7$ we
obtain the following  effective action,

\be \label{eq7l} \renewcommand{\arraystretch}{1.3}
\begin{array}{l} \langle Q_7 \rangle \Rightarrow  -
\frac{\dis 3}{\dis 32 \pi} \, f_\pi^2 \, \left[ 2 \, i \,  \langle
f^{\mu \nu}_{(+)} \, \left\{ \Delta_{3 2} \, , \, \xi_\mu
\xi_\nu  \right\} \rangle \, + \, 2\, i \, \langle f^{\mu
\nu}_{(-)} \,  \left\{ \Delta_{3 2} \, , \, \xi_\mu \xi_\nu
\right\} \rangle  \right. \\ \left.  \hspace*{0.5cm} - \, 2\,
\langle f_{(-)}^{\mu \nu} \, f_{(-) \mu \nu} \, \Delta_{3 2}
\rangle \, - \, \langle \left\{f_{(-)}^{\mu \nu}\, , \, f_{(+)
\mu \nu}\right\}
 \, \Delta_{3 2} \rangle \right] , \ea
\renewcommand{\arraystretch}{1} \ee

\vspace*{0.5cm}

\noindent which is valid to all orders in the
$1/N_c$-expansion.

The terms in Eqs. (\ref{eq6}) and (\ref{eq7l}) break the
relation  $\omega_1 \, = \, \omega_2$ which was found to
leading ${\cal O} (N_c^2)$.

\subsection{Results} \label{4.3} \indent

The coupling constants $g_8 \, \omega_{1,2,4}$ and $g_8 \,
\omega' _{1,2}$  in Eq. (\ref{effla}) can now be read off
from   the expression of the low-energy $\Delta S \, = \,
1$ Hamiltonian, by  inserting the effective action of the
four-quark operators $Q_-$, $Q_+$, $Q_3$, $Q_4$, $Q_5$,
$Q_6$ and the electroweak penguin operator
 $Q_7$ that are in Eqs. (\ref{eq-})-(\ref{eq5}) and
(\ref{eq6})-(\ref{eq7l}),  in Eqs. (\ref{hamil}) and
(\ref{gwh}). To evaluate these coupling constants, we have
fixed the matching  renormalization scale $\Lambda_\chi$
of the Wilson coefficients  and the effective action of
four-quark operators  to be of the order of the first
vector meson resonance mass,  ($\Lambda_\chi \, = \,
M_\rho \, = \, 770$ MeV).

We also need to know the coupling constant $g_8$ defined
in Eq. (\ref{eg8}).  This coupling, within this framework was
already computed in Ref. \cite{pr}. The result we get at
the $\Lambda_\chi$ scale is the following,

\be \label{g8} \renewcommand{\arraystretch}{1.3}
\begin{array}{l} g_8 \, = \, \frac{\dis 1}{\dis 2} \, C_-
(\Lambda_\chi) \, \left(1 \, - \,  \frac{\dis g_0}{\dis N_c} \,
- \, \gamma_-(\Lambda_\chi) \right) \, + \,  \frac{\dis
1}{\dis 10} \, C_+ (\Lambda_\chi) \,  \left(1 \, + \,
\frac{\dis g_0}{\dis N_c} \, - \, \gamma_+(\Lambda_\chi)
\right) \\  \hspace*{0.5cm}  + \, C_4 (\Lambda_\chi) \, + \,
\frac{\dis 2 \pi}{\dis N_c} \,  \alpha_s (\Lambda_\chi) \,
\left(2 \, H_1 \, + \, L_{10}  \right) \, \left( C_+
(\Lambda_\chi) \, + \, C_- (\Lambda_\chi) \right) \\
\hspace*{0.5cm} - \, 16 \, L_5 \, \frac{\dis \langle
\overline \Psi \Psi \rangle ^2} {\dis f_\pi^6} \, \left[ C_6
(\Lambda_\chi) \, + \,  \frac{\dis 2 \pi}{\dis N_c} \,
\alpha_s (\Lambda_\chi) \,  \left(2 \, H_1 \, + \, L_{10}
\right) \right. \\ \left.  \phantom{\frac{\dis \alpha_s
(\Lambda_\chi)}{\dis \pi}} \hspace*{4cm}  \times \, \left(
C_+ (\Lambda_\chi) \, + \, C_- (\Lambda_\chi) \right)
\right] \, + \, {\cal O} (\alpha_s N_c). \ea
\renewcommand{\arraystretch}{1} \ee

\vspace*{0.5cm}

\noindent We recall that $L_5$ is one of the ${\cal O}
(p^4)$ constants needed to renormalize the UV-behaviour
of the lowest order chiral loops \cite{gl1}. In the
large-$N_c$ limit $L_5$ is order $N_c$.

It turns out that the measurable quantities in the
transitions we are interested in only depend on the
combinations:  $g_8 (\omega_1 \, - \, \omega_2)$,  $g_8
(\omega_2 \, - \, 4 L_9)$,   $g_8 (\omega_4 \, - \, 6
L_{10})$ and $g_8 (\omega_1 \, + \, 2 \, \omega_2 \, - \,
\omega'_1 \, +  \, 2 \, \omega'_2)$, \cite{epr1,ep}. In the
large-$N_c$ limit   the combination $(\omega_1 \, - \,
\omega_2) / L_9$ is order $1/N_c$ whereas the
combinations $(\omega_2 \, - \, 4 \, L_9) / L_9$ and
$(\omega_4 \, - \, 6 \, L_{10}) / L_{10}$ are order $1$. From
our calculation we find that the difference $\omega_1 \, -
\, \omega_2$ depends only on the penguin operators
$Q_6$ and $Q_7$.

At this point, it is worth coming back to the question of
factorization of the coupling constant $g_8$ in the r.h.s of
Eq. (\ref{w1w2}). The expression for the effective action
of the Hamiltonian in Eqs.  (\ref{hamil}) and (\ref{gwh})
calculated at order $p^4$ in the chiral  expansion and at
order $N_c (\alpha_s N_c)$ in the $1/N_c$-expansion
together with the expression of  the $g_8$ coupling
constant in Eq. (\ref{g8}) lead us to the conclusion that the
approach of factorizing out $g_8$ in Eq. (\ref{w1w2}) is not
valid when one considers next-to-leading corrections in
the $1/N_c$-expansion. Therefore in the rest of the paper
we shall give results  for the combinations $g_8 \,
\omega_{1,2,4}$ and $g_8 \, \omega'_{1,2}$.

\section{Analysis of the results} \label{5} \indent

Let us now analyse some phenomenological implications
which follow from our  calculation of the coupling
constants  $g_8 \, \omega_{1,2,4}$ and $g_8 \,
\omega_{1,2}'$. In Ref. \cite{epr1}, the
 decay amplitudes of $K \to \pi \gamma^*$ were
calculated at the one-loop
 level with the following results:

\be \ba A(K^+ \to \pi^+ \gamma^*) \, = \, \frac{\dis
G_F}{\dis \sqrt 2}\,  V_{ud} V_{us}^* \, \frac{\dis g_8}{\dis
16 \pi^2} \, q^2\, \widehat \Phi_+ (q^2) \, \epsilon^\mu
(p+p')_\mu \,; \ea \ee

\be \ba A(K^0_S \to \pi^0 \gamma^*) \, = \, \frac{\dis
G_F}{\dis \sqrt 2}\,  V_{ud} V_{us}^* \, \frac{\dis g_8}{\dis
16 \pi^2} \, q^2\, \widehat \Phi_S (q^2) \, \epsilon^\mu
(p+p')_\mu \,; \ea \ee

\noindent with

\be \label{42} \renewcommand{\arraystretch}{1.3} \ba
\widehat \Phi_+ (q^2) \, = \, \frac{\dis 16 \pi^2}{\dis 3} \,
 \left[ (\omega_1^r \, - \,  \omega_2^r) \, + \,
3(\omega_2^r \, - \, 4 L_9^r) \right] (\nu^2) \\
\hspace*{1.5cm} - \,  \left[ \Phi_K (q^2) \, + \,
\Phi_\pi(q^2) \, - \, \frac{\dis 1}{\dis 3} \ln \left(
\frac{\dis m_\pi m_K}{\dis \nu^2} \right) \right]; \ea
\renewcommand{\arraystretch}{1} \ee

\be \ba \widehat \Phi_S (q^2)\, = \, -\, \frac{\dis 16
\pi^2}{\dis 3} \,  \left( \omega_1^r \, - \, \omega_2^r
\right)(\nu^2) \,  + \, 2 \, \Phi_K (q^2) - \, \frac{\dis 1}{\dis
3} \ln \left( \frac{\dis m_K^2}{\dis \nu^2} \right). \ea \ee

\noindent Here,

\be \renewcommand{\arraystretch}{1.3} \ba \Phi_{K (\pi)}
(q^2) \, = \, - \frac{\dis 4 m_{K (\pi)}^2}{\dis 3 q^2}\, +  \,
\frac{\dis 5}{\dis 18} \, + \, \frac{\dis 1}{\dis 3}
\left(\frac{\dis 4 m_{K (\pi)}^2}{\dis q^2} \, - \, 1
\right)^{3/2}\, \arctan \left( 1/
 \sqrt {\frac{\dis 4 m_{K (\pi)}^2}{\dis q^2}\, - \, 1} \right)
\\ {\rm for} \, q^2 \le 4 m_{K (\pi)}^2. \ea
\renewcommand{\arraystretch}{1} \ee

\noindent The coupling constants $\omega_1$,
$\omega_2$, $\omega_4$, $\omega' _1$ and   $\omega '_2$
are scale dependent quantities. In our approach,  the
explicit scale dependence of $\omega_{1,2,4}$ and
$\omega_{1,2}'$  comes from next-to-leading terms which
we have not calculated. We identify the values we get for
$\omega_{1,2,4}$ and $\omega '_{1,2}$ with those of
$\omega_{1,2,4}^r$ and $\omega _{1,2}^{'r}$  renormalized
at the $\rho$-resonance mass. We shall also identify the
constant
 $L_9^r$ in Eq. (\ref{42}) with the coupling $L_9$
renormalized at this same
 scale. Following Ref. \cite{epr1} we define the constants,

\be \label{w+} \ba \omega_+ \equiv  - \frac{\dis 16
\pi^2}{\dis 3} \, \left[ (\omega_1^r \, - \, \omega_2^r) \, +
\, 3 ( \omega_2^r \, - \, 4 L_9^r) \right] (M_\rho^2) \, - \,
\frac{\dis 1}{\dis 3} \ln \left(\frac{\dis m_K m_\pi} {\dis
M_\rho^2} \right); \ea \ee

\be \label{ws} \ba \omega_S \equiv  - \frac{\dis 16
\pi^2}{\dis 3} \,  \left( \omega_1^r \, - \, \omega_2^r
\right) (M_\rho^2) \, - \, \frac{\dis 1}{\dis 3}
 \ln \left( \frac{\dis m_K^2}{\dis M_\rho^2}\right). \ea \ee

\noindent Thus

\be \ba \widehat \Phi_+(q^2) \, = \, - \left[\Phi_K(q^2) \, +
\, \Phi_\pi(q^2) \, + \, \omega_+\right]; \ea \ee

\be \ba \widehat \Phi_S(q^2) \, = \, 2 \Phi_K(q^2) \, + \,
\omega_S; \ea \ee

\noindent and the decay rates for $K \to \pi l^+ l^-$ can be
written in the following way

\be \label{decay} \ba \Gamma (K \to \pi l^+ l^-) \, = \,
\overline \Gamma {\dis \int_{4\epsilon}^ {(1-\sqrt
\delta)^2}} d z \, \lambda^{3/2}(1,z,\delta) \, \left(1\,-\,4
\frac{\dis \epsilon}{\dis z}\right)^{1/2}\, \left(1\, + \,
2\frac{\dis \epsilon} {\dis z}\right) |\widehat \Phi|^2, \ea
\ee

\noindent where

\be \renewcommand{\arraystretch}{1.5} \ba z\, =\,
\frac{\dis q^2}{\dis m_K^2}, \, \epsilon \, = \, \frac{\dis
m_l^2}{\dis m_K^2}, \, \delta \, = \, \frac{\dis
m_\pi^2}{\dis m_K^2},\\ \lambda(x,y,z) \, = \,
x^2+y^2+z^2-2xy-2yz-2zx,  \ea
\renewcommand{\arraystretch}{1} \ee

\noindent and $\overline\Gamma$ is an overall
normalization factor,

\be \ba \overline\Gamma \, = \, \left| \frac{\dis G_F}{\dis
\sqrt 2} \, V_{ud}  V_{us}^* \right|^2 \, \frac{\dis \alpha^2
m_K^5 |g_8|^2}{\dis 12 \pi  (4 \pi)^4}.  \ea \ee

\noindent In our numerical estimates we shall  use the
following set of input values:

\be \renewcommand{\arraystretch}{1.3} \ba \langle
\widehat{\bar q q} \rangle \, = \, - ((190-210) \, {\rm
MeV})^3;  \,\,\,  \langle \frac{\dis \alpha_s}{\dis \pi} G^2
\rangle   \, = \, ((330-390) \, {\rm MeV})^4;  \\
\Lambda_\chi \, = \, (700-900) \, {\rm MeV}; \,\,\,
\Lambda_{\overline{MS}} \, = \, (100-200) \,\,  {\rm MeV} \\
{\rm and} \\  M_Q \, = \, (250-350) \, {\rm MeV}. \ea
\renewcommand{\arraystretch}{1} \ee

\noindent

\noindent Then we have the following result for $g_8 {\rm Re} \,
\omega_+$, $g_8 {\rm Re} \, \omega_S$, $g_8 \, (\omega_1 \, + \, 2 \,
\omega_2)$ and $g_8 \, (\omega_1 \, - \, \omega_2)$,

\be \label{cal} \renewcommand{\arraystretch}{1.3} \ba
g_8 {\rm Re} \, \omega_+ \, = \, 7.5^{+5}_{-3}; \\ g_8{\rm Re} \,
\omega_S \, = \,  5^{+4}_{-2}; \\ g_8 \, (\omega_1 \, + \, 2 \,
\omega_2) \, = \, 0.12 ^{+0.02}_{-0.01}; \\ g_8 \, (\omega_1 \, - \,
\omega_2) \, = \, - \, 0.08 ^{+0.04}_{-0.08},  \ea
\renewcommand{\arraystretch}{1} \ee

\noindent where the central value corresponds to the
input values  $\langle \widehat {\bar q q} \rangle \, = \, - (
200 \, {\rm MeV})^3$, $\langle \frac{\dis \alpha_s}{\dis \pi}
G^2 \rangle  \, = \, (360 \, {\rm MeV})^4$,
$\Lambda_{\overline{MS}} \, = \, 150$ MeV and $M_Q \, = \,
300$ MeV with $\Lambda_\chi \, = \, 800$ MeV.
Experimentally \cite{bnl} we know that,

\be \label{exp} \renewcommand{\arraystretch}{1.3} \ba
g_8 {\rm Re} \, \omega_+\, = \, 4.6 ^{+ 1.2} _{- 0.7}; \\ g_8
\, (\omega_1 \, + \, 2 \, \omega_2) \, = \, 0.41 ^{+ 0.10}_{-
0.05}. \ea \renewcommand{\arraystretch}{1} \ee

\vspace*{0.5cm}

\noindent In view of the these experimental results a value of the
quark condensate lower than $- (210 \, {\rm MeV})^3$
turn out to be not favoured.

Our results tell us that the combination of counterterms
for   $K^+ \to \pi^+ \, l^+ l^-$, $K^0 \to \pi^0 \, l^+ l^-$ and
$\eta \to \bar K^0 \, l^+ l^-$ decay amplitudes,
$\omega_S$ and $\omega_+$  \cite{epr1}, depend strongly
on the penguin diagrams (both hadronic and  electroweak).
In addition, it turns out that the coupling  $g_8 \,
\omega_S$ only depends on the penguin operators
whereas  $g_8 \, \omega_+$ depends also on the
non-penguin Wilson coefficients.  This fact could be used
for measuring their respective strength.

With the value of $\omega_S$ in Eq. (\ref{cal}) we can
predict the  following branching ratio,

\be \label{bra} \renewcommand{\arraystretch}{1.5}
\begin{array}{l} \Gamma (K^0_S \to \pi^0 e^+ e^-) \simeq
2.6 \times 10^{-9} \, \left[ {\rm Re} \, \omega_S^2 \, - \,
0.66 \, {\rm Re} \, \omega_S \, +  \, 0.11 \right] \\
\hspace*{3.2cm} \simeq (0.5\,- \, 5) \times 10^{-9} , \ea
\renewcommand{\arraystretch}{1} \ee

\vspace*{0.5cm}

\noindent for which there is an experimental upper bound
of the order of $10^{-5}$ \cite{pdg}. We can also give a
prediction for the ratio of decay  rates of  the $K^+ \to
\pi^+ e^+ e^-$ and $K^0_S \to \pi^0 e^+ e^-$ transitions;

\be \label{ratio} \ba \frac{\dis \Gamma(K^0_S \to \pi^0
e^+ e^-)}
     {\dis \Gamma(K^+ \to \pi^+ e^+ e^-)} \simeq   \frac{\dis
{\rm Re} \, \omega_S^2 \, - \, 0.66 \, {\rm Re} \, \omega_S
\, + \, 0.11}{\dis {\rm Re}\, \omega_+^2 \, - \, 0.59 \, {\rm
Re} \, \omega_+
 \, + \, 0.09} \, = \, 0.30^{+0.50}_{-0.25}. \ea \ee

\vspace*{0.5cm}

The coupling constant $\hat c$ introduced in Eq. (\ref{hc})
can also be determined from the results above. It turns
out that there is no contribution {}from the electroweak
penguin $Q_7$ to the $\hat c$ coupling constant.  For its
real part we find

\be {\rm Re} \, \hat c \, = \, - \, 0.7 \pm 0.5.  \ee

\vspace*{0.5cm}

\noindent which translates in the following  prediction for
the branching ratio of the transition  $K^+ \to \pi^+
\gamma \gamma$ \cite{epr1}:

\be \label{kpgg} \ba {\rm BR} (K^+ \to \pi^+ \gamma
\gamma) \, = \, (5.2 \pm 0.7) \times 10^{-7} . \ea \ee

\vspace*{0.5cm}

\noindent The experimental upper limit depends very
much on the $\pi^+$  energy spectrum, giving a wide range
of allowed values \cite{bnl1},

\be \ba {\rm BR} (K^+ \to \pi^+ \gamma \gamma) \, \le \,
1.5 \times 10^{-4} . \ea \ee

\vspace*{0.5cm}

\noindent The imaginary part of $\hat c$ vanishes since
there is no contribution of the electromagnetic penguin
$C_7^\gamma$ to this coupling. This implies that there is
no  charge asymmetry  $\Gamma (K^+ \to \pi^+ \gamma
\gamma) \, -
 \, \Gamma (K^- \to \pi^- \gamma \gamma)$ from the
CP-violating phase of the CKM mixing matrix \cite{epr1}
appearing in the electromagnetic penguin Wilson
coefficient.

Finally, for the combinations $\omega_1 \, + \, 2\,
\omega_2 \, - \,  \omega_1'$ and $\omega_2'$ introduced
in Eq. (\ref{effla}) we get the  results,

\be \label{an} \renewcommand{\arraystretch}{1.5} \ba g_8
\, ( \, \omega_1 \, + \, 2 \, \omega_2 \, - \, \omega_1') \, =
\, 0.02 \pm 0.01 \, , \\ \omega_2' \, = \, 0 \, . \ea
\renewcommand{\arraystretch}{1} \ee

\vspace*{0.5cm}

\noindent These two combinations turn out to be
independent of the electroweak penguin operator $Q_7$
and therefore real. They can be used in the theoretical
prediction of the electric-type amplitude of the $K^+ \to
\pi^+ \pi^0 \gamma$ transitions, see Ref. \cite{ep}.

\section{Conclusions} \label{6} \indent

In the framework of the effective action aproach for
four-quark operators \cite{pr},  we have
calculated the various coupling constants that enter in
the chiral perturbation theory prediction for the   $K \to
\pi \gamma^* \to \pi l^+ l^-$, $K \to \pi \gamma \gamma$
and  $K \to \pi \pi \gamma$ decay rates, \cite{epr1,ep}.
These constants are not  determined by symmetry
requirements alone. They turn out to depend strongly on
 the effective action of the penguin operators $Q_6$ and
$Q_7$.

We have given a prediction for the phenomenological
constants  $\omega_+$, $\omega_S$ and $\hat c$ defined
in Ref. \cite{epr1}, which fix  the decay rates for $K \to \pi
\gamma^*$ and $K^+ \to \pi^- \gamma \gamma$.  In
particular we have predicted the branching ratio for
$K^0_S \to \pi^0 e^+ e^-$ in (\ref{bra}) and the ratio of
decay rates of  $K^0_S \to \pi^0 e^+ e^-$ and $K^+ \to
\pi^+ e^+ e^-$ in (\ref{ratio}). We
 have given the branching ratio for $K^+ \to \pi^+ \gamma
\gamma$ in  (\ref{kpgg}) and found that there is no charge
asymmetry  $\Gamma (K^+ \to \pi^+ \gamma \gamma) \, -
\, \Gamma (K^- \to \pi^- \gamma \gamma)$  coming from
the electromagnetic  penguin Wilson coefficient
$C_7^\gamma$. In the transition  $K \to \pi \pi \gamma$,
further counterterms are possible. They have been
classified in Ref. \cite{ep}. The coupling constants that
modulate these new  counterterms $\omega '_1$ and
$\omega '_2$ have also been predicted in  (\ref{an}).

\section*{Acknowledgements} \indent

We wish to thank Eduardo de Rafael for suggesting this
calculation to us and for many useful discussions. We have
also benefited from discussions with Gerhard Ecker, Toni
Pich and Josep Taron. We would like to thank Lars
H\"ornfeldt for his help with the algebraic manipulating
program STENSOR. The work of one of us (J.P.) has been
supported in part by CICYT, Spain, under Grant No.
AEN90-0040. J.P. is also indebted to the Spanish
Ministerio de  Educaci\'on y Ciencia for a fellowship.

 \end{document}